\documentclass[letter,12pt,authoryear]{elsarticle}

\usepackage{amssymb}
\usepackage{amsmath}

\usepackage{graphicx}        
\newcommand{\be}{\begin{equation}}
\newcommand{\ee}{\end{equation}}
\newcommand{\bra}{\langle}
\newcommand{\ket}{\rangle}
\newcommand{\bea}{\begin{eqnarray}}
\newcommand{\eea}{\end{eqnarray}}

\begin{document}

\begin{frontmatter}



\title{Multifractality and sample size influence on Bitcoin volatility patterns
}


\author{Tetsuya Takaishi} 
\ead{tt-taka@hue.ac.jp}
\affiliation{organization={Hiroshima University of Economics},
            addressline={}, 
            city={Hiroshima},
            postcode={731-0192}, 
            state={},
            country={JAPAN}}

\begin{abstract}
The finite sample effect on the Hurst exponent (HE) of realized volatility time series
is examined using Bitcoin data.
This study finds that the HE decreases as the sampling period $\Delta$ increases and a simple finite sample ansatz closely fits the HE data.
We obtain values of the HE as $\Delta \rightarrow 0$, which are smaller than 1/2, indicating rough volatility.
The relative error is found to be $1\%$ for the widely used five-minute realized volatility. 
Performing a multifractal analysis, we find the multifractality in the realized volatility time series, smaller than that of the price-return time series.

\end{abstract}


\end{frontmatter}



\section{Introduction}
Volatility as a measure of risk is of great importance in empirical finance, especially in the risk management sector.
Forecasting future volatility is an essential task for financial institutions and practitioners
to manage financial assets safely and avoid unacceptable losses in the future. 
Practically forecasting volatility is implemented by assuming models that
mimic the properties of price dynamics.
It is well-known that universal properties exist across various assets, denoted as
"stylized facts"~\citep{Cont2001QF}.
One of main universal properties is "volatility clustering".
\cite{Engle1982autoregressive} introduced the autoregressive conditional heteroscedasticity (ARCH) model
that captures the property of volatility clustering. Later the ARCH model was generalized to
GARCH model by \cite{Bollerslev1986JOE}.  
Another notable property is that volatility is long-correlated, which is also related to 
the fact that the absolute return $|r|$ (as a proxy of volatility) 
displays long autocorrelations \citep{ding1993long}.
A more precise measure of volatility, realized volatility (RV)~\citep{andersen1998answering}, shows long-memory behavior \citep{andersen2001exchange}. 
According to these empirical findings, it is natural to include  long-memory characteristics in volatility modelings.
Since long-memory time series are characterized by the Hurst exponent (HE) of $H>1/2$, 
the fractional volatility model (FVM) 
is introduced  using  the fractional Brownian motion with $H>1/2$\citep{comte1998long}.

A new paradigm on volatility dynamics suggests 
that "volatility is rough." \citet{gatheral2018volatility} 
argue that the $H$ of RV time series is  approximately 0.1, less than 1/2.
This observation implies that volatility time series exhibit rough or anti-persistent behavior.
Considering this observation, they suggest the rough FVM (RFVM) with $H<1/2$ and show that the RFVM 
improves volatility forecasts.
Furthermore, several advantages to use rough volatility are indicated.
One is the volatility surface issue. In particular, it is empirically observed that the term structure of at-the-money skew is described by a negative power law, which is not easily explained 
by conventional stochastic volatility models\citep{carr2003finite,fouque2003multiscale,lee2005implied}.
To this issue, it is shown that models based on fractional Brownian motion are capable of explaining
the negative power law\citep{alos2007short,fukasawa2011asymptotic}.
Another issue is the Zumbach effect\citep{zumbach2003volatility,zumbach2009time} implying that
the cross-correlation function between the daily variance and squared returns 
has the time-reversal asymmetry which is not derived from conventional volatility models.
Interestingly, \citet{el2020zumbach} shows that the rough Heston model could explain 
the Zumbach effect.
Since there exist several issues that can be explained by the rough volatility,
it can be interesting to further seek evidence solved by rough volatility.
There also exist further developments to use rough volatility models for 
such as option pricing\citep{bayer2016pricing} and perfect hedging\citep{euch2018perfect} and
investigations of origins of roughness by the market microstructure\citep{el2018microstructural,jusselin2020no,rosenbaum2021microscopic}.

Various empirical studies, using data from the RVs of various assets, imply that volatilities confirm the roughness of volatility \citep{bennedsen2022decoupling,livieri2018rough,takaishi2020rough,floc2022roughness}.
Contrary to these empirical findings,
the roughness of volatility remains a  controversial issue. 
For example, \citet{cont2024rough} show that 
the HE is different from spot and integrated (realized) volatilities and
infer that this may arise due to measurement errors of RV. 
\citet{Fukasawa2019IsVR} find that the RV is still rough using an improved estimator;  even more so than the observations of previous researchers.
\citet{brandi2022multiscaling} argue that the rough Bergomi model proposed for rough volatility is inconsistent with multi-scaling properties.

In this study, we address the finite sample effect on the HE obtained from RV.
RV is constructed by summing the squared returns sampled at a certain frequency; 
generally, the number of samples to construct the RV is finite.
When the number of samples is finite, the RV receives the finite sample effect and, consequently, 
the distribution of returns standardized by the RV deviates from a Gaussian distribution \citep{peters2006testing}.
Owing to the absence of previous research on this topic, we examine impact of the finite size effect on the HE estimation of RV time series.

The HE relates to the scaling behaviour of 2nd order fluctuations or variance.
For the random time series, $H=1/2$.
For $H>1/2 (<1/2)$, the time series is said to be persistent (ant-persitent).
The HE can be generalized for q-th order fluctuations and 
it is referred to as the generalized Hurst exponet (GHE) $h(q)$.
When $h(q)$ is constant for any $q$, such time series 
is said to be monofractal.
Conversely, when $h(q)$ varies for $q$, it is multifractal.
The GHE can capture the non-linear time-correlations that can not be 
measured in the HE alone.
Numerous studies on the GHE have been conducted for price-return time series 
and it is found that the multifractal nature is usually present in price-return time series, e.g. see \cite{Jiang-Xie-Zhou-Sornette-2019-RPP}. 

In the literature, the RV time series is considered to be monofractal \citep{gatheral2018volatility}
and there is little attention in multifractality.
However, the possibility of multifractality in the RV 
is pointed out by \citet{takaishi2020rough}. 
Small multifractality  in the RV of stock returns is also observed~\citep{brandi2022multiscaling}.
In this study, we perform a multifractal analysis for the RV time series
to clarify the existence of the multifractality in the RV.

\section{Data and Methodology \label{sec2}}

Our data consists of Bitcoin tick data traded on Bitstamp exchange
from January 2, 2014, to June 01, 2023.
We do not use data prior to 2014 because, in the early stages of the Bitcoin market, 
liquidity 
was low \citep{takaishi2019market}, 
and we can observe different market properties in ill-liquid markets.
The HE of the return time series in the early stages of the Bitcoin market
was less than 1/2, indicating the anti-persistence of the time series\citep{urquhart2016inefficiency}.
Subsequently, the HE increased to 1/2 as the liquidity increased~\citep{takaishi2019market}.
It is argued that the anti-persistence of returns seen in the cryptocurrency market can be attributed to the low liquidity of the market~\citep{wei2018liquidity}.

We construct the daily RV on day $t$ 
with $\Delta$-minute period by 
\be
RV_{t,\Delta} = \sum_i^{n} r_{t,i,\Delta}^2,
\ee
where  $r_{t,i,\Delta}, i=1,2,\dots, n$ are intraday returns
and $n=1440/\Delta$ is the number of samples in one day.

At finite $\Delta$, the RV receives the finite sample effect that could lower the accuracy of the RV estimate.
Let us assume that the observed daily return $r_t$ is described by $r_t=\sigma_t\epsilon_t$,
where $\epsilon_t\sim N(0,1)$ and $\sigma_t$ is the standard deviation.
Under this assumption, the distribution of $r_t/\sigma_t$ should be the standard normal distribution.
Using the RV as a proxy of $\sigma_t$,
\citet{andersen2000exchange} show that the distribution of $\bar{r_t}\equiv r_t/RV_t^{1/2}$ is nearly Gaussian.
However, when  $n$ is small,
the distribution of $\bar{r_t}$ deviates from the Gaussian distribution. 
\citet{peters2006testing} provide the finite sample formula of 
the probability distribution $P(\bar{r_t})$:
\be 
P(\bar{r})=\frac{\Gamma(n/2)}{\sqrt{\pi}n\Gamma((n-1)/2)} \left(1-\frac{\bar{r}^2}{n}\right)^{(n-3)/2}.
\label{eq:p}
\ee
Empirical observations confirm
Eq.(\ref{eq:p}) \citep{takaishi2012finite}.
The 2k-th order moments of the standardized returns are also affected by the finite sample effect as
\be
E[\bar{r}^{2k}]=\frac{n^k(2k-1)(2k-3)\dots 1}{(n+2k-2)(n+2k-4)\dots n}.
\label{eq:moments}
\ee
From Eq.(\ref{eq:moments}), the kurtosis at finite $n$ is given by
\be
\frac{E[\bar{r}^{4}]}{E[\bar{r}^{2}]^2}=\frac{3n}{n+2}.
\label{eq:kurtosis}
\ee

The RV time series is defined by log-volatility increments as follows.
\be
V_t = \log RV_t - \log RV_{t-1}.
\ee 
We determine the HE ($=h(2)$) of the time series $V_t$
using the multifractal detrended fluctuation analysis (MDFA), suitable for 
non-stationary processes \citep{kantelhardt2002multifractal}
and widely used for investigations of time series properties \citep{Jiang-Xie-Zhou-Sornette-2019-RPP}.
The MFDFA is described as follows. For a more detailed description, see, e.g., \citet{kantelhardt2002multifractal}.

First, we determine the profile $Y(i)$ from $V_t$
\be
Y(i)=\sum_{j=1}^i (V_j- \bra V \ket),
\ee
where $\bra V \ket$ stands for the average of $V_t$.
Then, we divide the profile $Y(i)$  into $N_s$ non-overlapping segments of an equal length $s$, where $N_s \equiv {int} (N/s)$.
Since the length of the time series is not always a multiple of $s$,
we repeat the same procedure, starting from the end of the profile.
Next, we calculate the variance $F^2$.
\be
F^2(\nu,s)=\frac1s\sum_{i=1}^s (Y[(\nu-1)s+i] -P_\nu (i))^2,
\ee
for each segment $\nu, \nu=1,\dots,N_s$ and
\be
F^2(\nu,s)=\frac1s\sum_{i=1}^s (Y[N-(\nu-N_s)s+i] -P_\nu (i))^2,
\ee
for each segment $\nu, \nu=N_s+1,\dots,2N_s$.
$P_\nu (i)$ is the fitting polynomial to remove the local trend in segment $\nu$.
Averaging all segments, we obtain the $q$th order fluctuation function
\be
F_q(s)=\left\{\frac1{2N_s} \sum_{\nu=1}^{2N_s} (F^2(\nu,s))^{q/2}\right\}^{1/q}.
\label{eq:FL}
\ee
If the time series is long-range power-law correlated,
$F_q(s)$ is expected to be the following functional form for large $s$:
\be
F_q(s) \sim s^{h(q)}.
\label{eq:asympto}
\ee
$h(q)$ is called the GHE and $h(2)$  corresponds to the HE.

\section{Results}

We determine the HE in eight-year window data using the rolling window method.
The eight-year window is rolled every five days;
in each window, we determine the HE by the MDFA
for the RV sampled at $\Delta =1,2,\dots 1440$, where $\Delta$ is chosen 
so that $1440/\Delta$ becomes an integer.

\begin{figure}
\centering
\includegraphics[height=7cm]{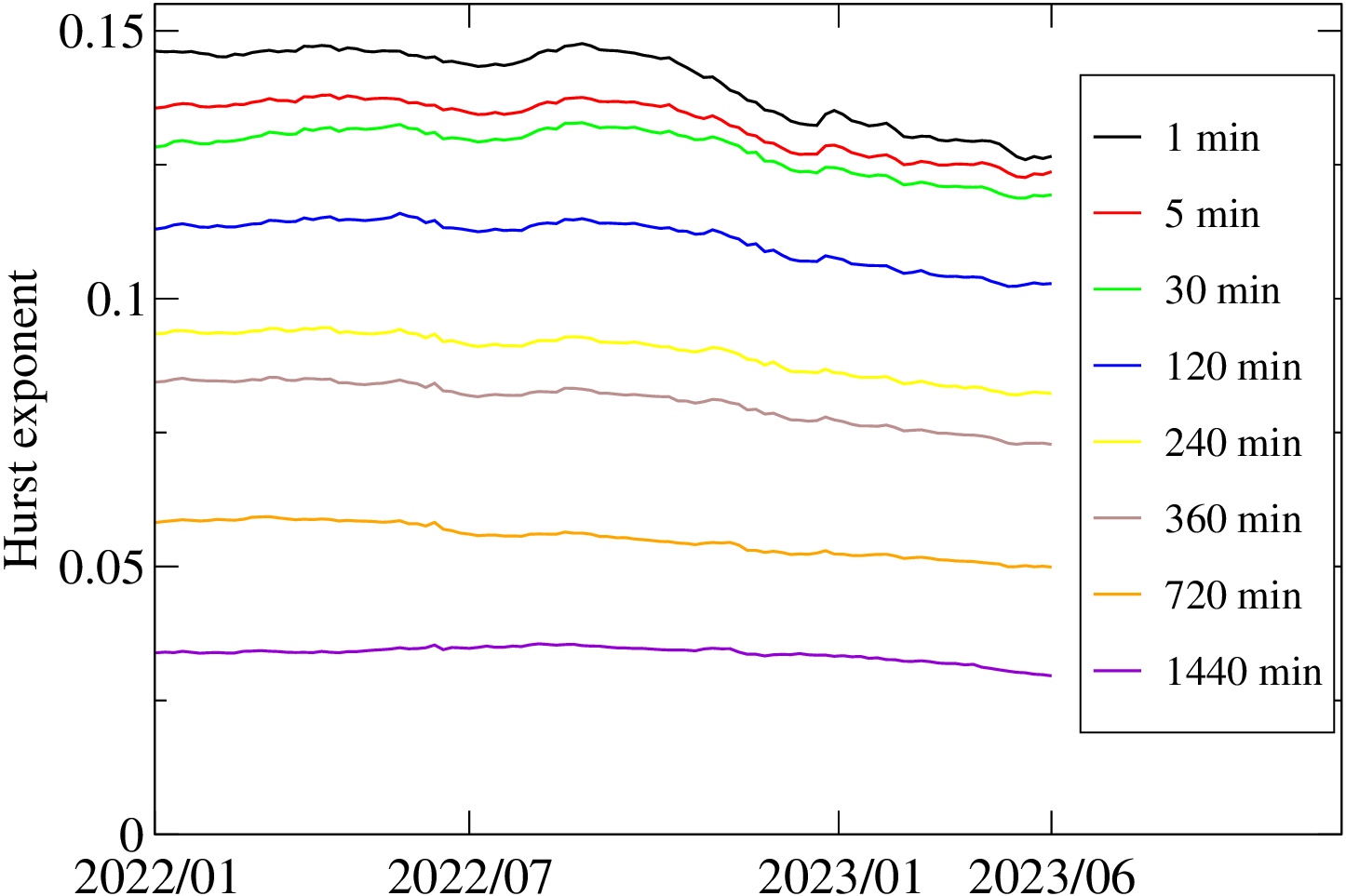}
\caption{
Time evolution of the HE ($h(2)$) for various sampling frequencies.
\label{fig.1}}
\end{figure}

Fig.\ref{fig.1} displays the time evolution of the HE for various $\Delta$.
We identify how HE tends to decrease as $\Delta$ increases.
Such decreasing behavior is also observed in \citep{garcin2022long}'s study of exchange rates.

\begin{figure}
\centering
\includegraphics[height=12cm]{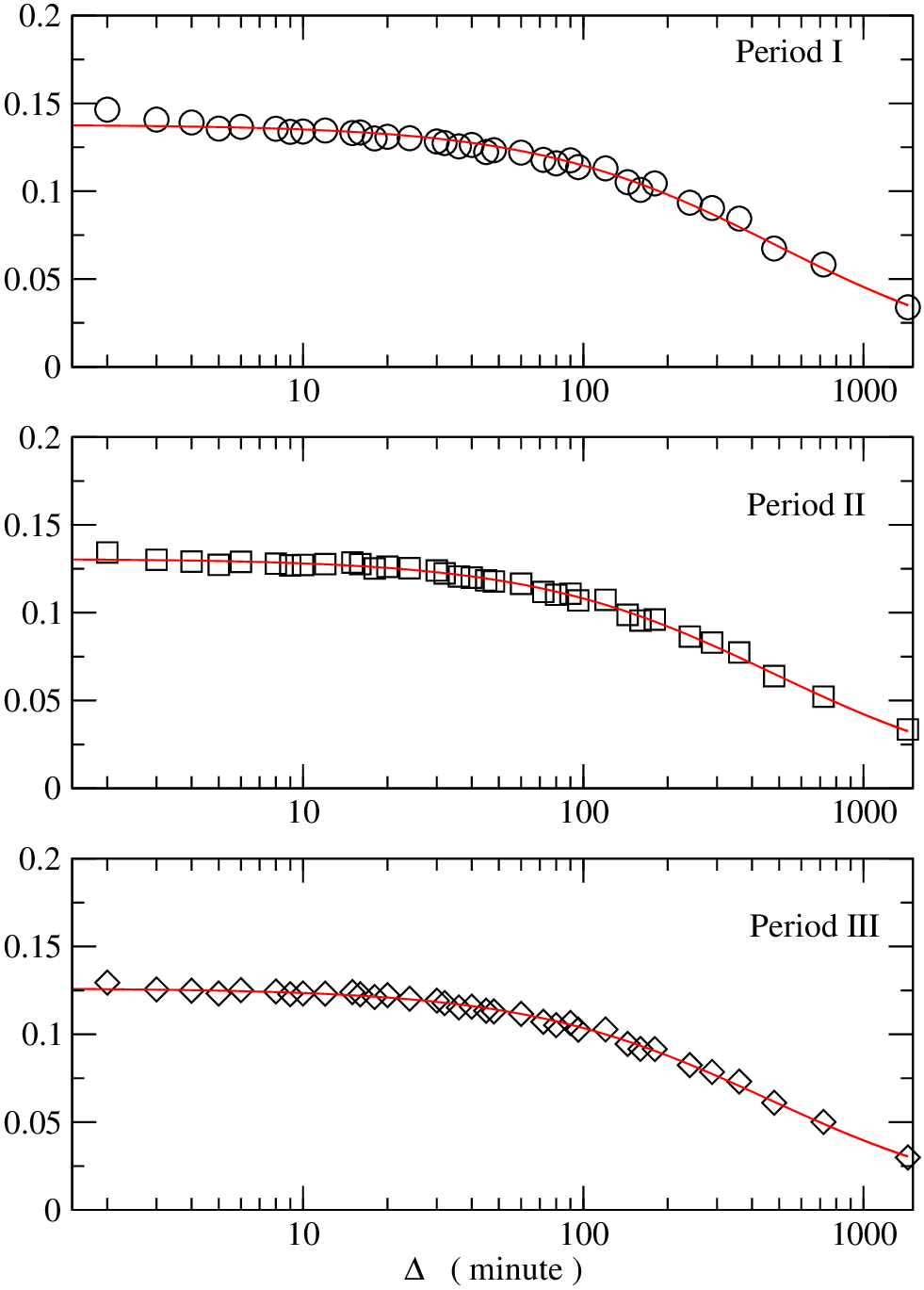}
\caption{
HE as a function of $\Delta$. Red solid lines show fitting results.
\label{fig.2}}
\end{figure}

To investigate the frequency dependence of the HE,
we select three representative data periods
(see Table 1) 
and plot the HE as a function of $\Delta$ in Fig.\ref{fig.2},
indicating that the HE decreases as $\Delta$ increases.
Although we do not know the finite sample formula for the HE,
inspired by the form of Eq.(\ref{eq:moments}),
we examine the following ansatz and find that the ansatz fits the HE results well.
\be
H(\Delta)=H_0 \frac{n}{n+a},
\label{eq:fit}
\ee
where $H_0$ and $a$ are fitting parameters.

The fitting results (see also Table 1 for the fitting parameters) are shown as red lines in Fig.\ref{fig.2},
indicating that 
Eq.(\ref{eq:fit}) fits the HE data well, 
except for the slight deviation at $\Delta$ = 1-minute.
The deviation at 1-minute could be related to the microstructure noise that
manifests at very high frequencies. 
\citet{fleming2003economic} finds that the kurtosis of standardized returns from US stocks exhibits a divergent behavior 
at very high frequencies and state that the divergent is caused by the leptokurtic distribution as a mixture of normals originated from the microstructure noise~\citep{hansen2006realized,alexander2001option}.
However, the microstructure noise effect in Bitcoin data that we observe seems small until about $\Delta=$1-min.
When the microstructure noise considerably affects the HE results, 
we need to fit the HE data while excluding such affects.

\begin{figure}
\centering
\includegraphics[height=7cm]{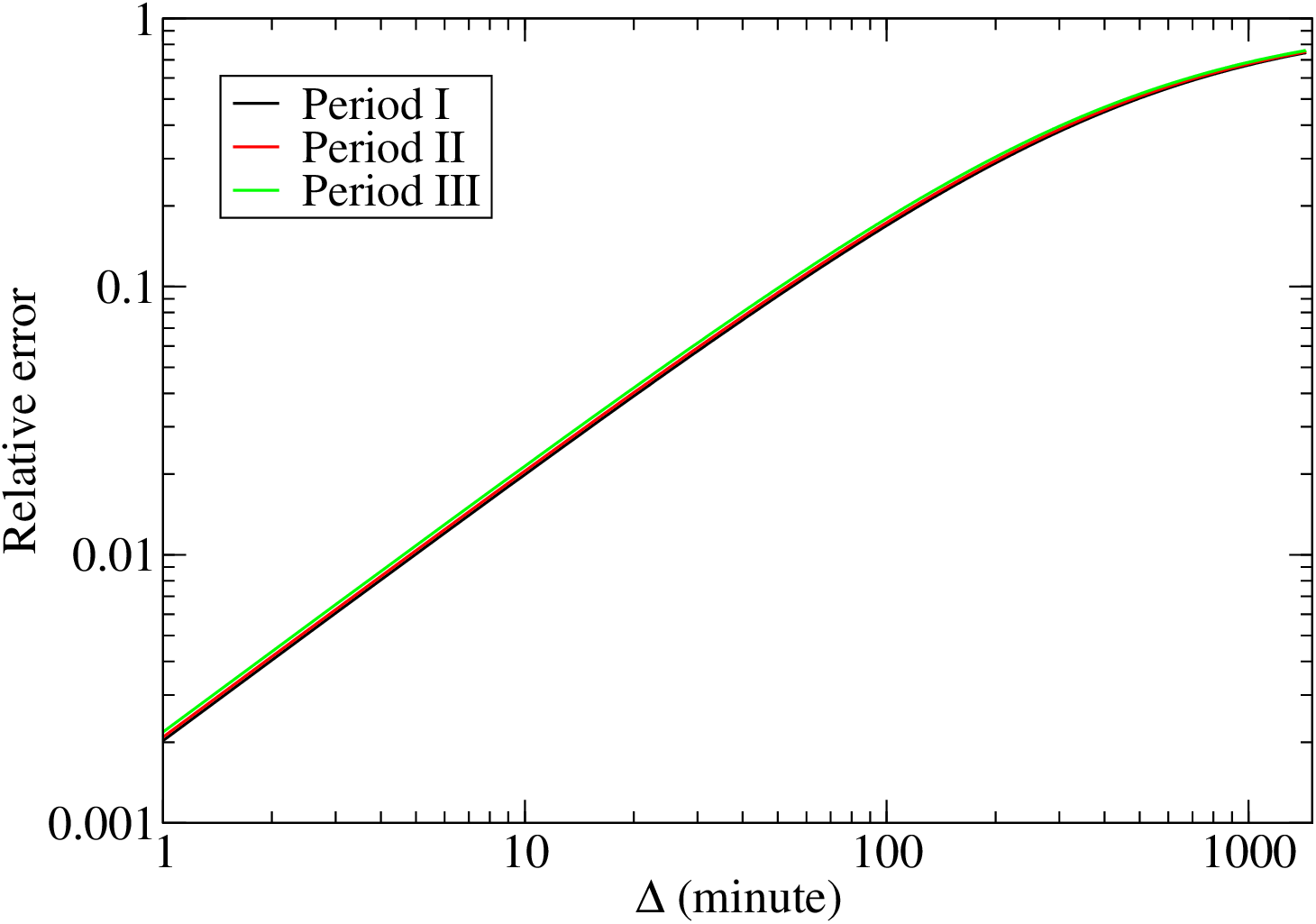}
\caption{
Relative errors to $H_0$ as a function of $\Delta$.
\label{fig.3}}
\end{figure}

The parameter $H_0$ corresponds to $H$ at $\Delta \rightarrow 0$. The obtained values of $H_0$
listed in Table 1 are around $0.12-0.14$ and exhibit a slight decreasing behavior as a function of time.
It is interesting to see that the parameter $a$ is close to an integer 3 as similar equations in Eq.(\ref{eq:moments})
although there is no reason that $a$ should be an integer.
The relative error to $H_0$ at finite $\Delta$ is given by $a/(n+a)$,
shown in Fig.\ref{fig.3}.
Three lines from different periods are very similar and difficult to distinguish visually. 
In empirical analyses, the sampling frequency for the RV is often $\Delta=5$-minute
since the five-minute RV gives a reasonable balance between the bias and the efficiency\citep{andersen1997intraday} 
and has better performance than other RV estimates \citep{liu2015does}.
The relative errors at five-minute  $\Delta$  are found to be around $1\%$, 
suggesting that the finite size effect on the HE  for the widely used five-minute RV is
reasonably small . 
Hereafter, we use the five-minute RV for the MFDFA.

Fig.4(a) shows the time evolution of $h(q)$
for $q=-3,2,3$, obtained using the five-minute RV time series.
It indicates that $h(q)$ as a function of $q$ is not constant, suggesting the multifractality of the time series.

To quantify the strength of multifractality, we define the following quantity\citep{zunino2008multifractal}:
\be 
\Delta h(k) = h(-k)-h(k),
\ee 
which goes to zero for the monofractal time series.
We also define the strength of multifractality by the Taylor coefficient of $h(q)$.
The function $h(q)$ is approximated linearly around $q=0$ as
\be 
h(q)=B_0+ B_1 q,
\ee 
where $B_0$ and $B_1$ are Taylor coefficients and the strength of multifractality 
is measured by $B_1$, which also takes zero for the monofractal time series.
Here, we approximately obtain $B_1$ by $-\Delta h(3)/6$.

Fig.4(b) and (c) display $\Delta h(3)$ and $-B_1$ as a function of time and
indicate that the strength of multifractality is finite and time-varying,
meaning the existence of multifractality in the RV time series.
The average value of $\Delta h(3)$ for whole period is 0.034.
We also perform the multifractal analysis for the price-return time series
and obtain $\Delta h(3)\simeq 0.12$, indicating a stronger multifractality than that of the RV time series.
Similarly, the average value of $-B_1$ for whole period is obtained to be 0.0057,
which is consistent with the similar strength of $B_1$ obtained for the stock RV time series\citep{brandi2022multiscaling}.

\begin{figure}
\centering
\includegraphics[height=12cm]{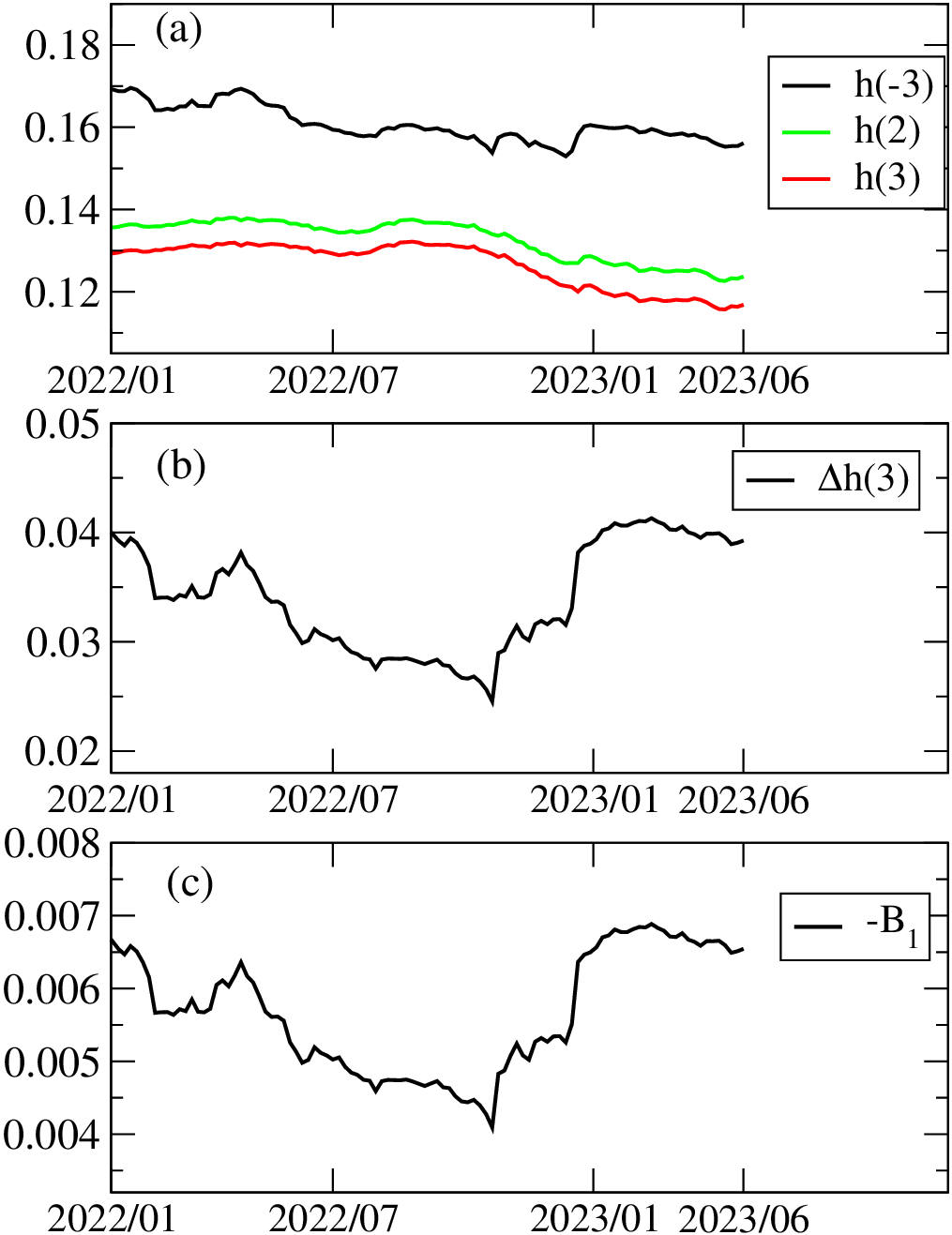}
\caption{
(a) Time evolution of $h(-3),h(2)$ and $h(3)$.
(b) Multifractal strength $\Delta h(3)$.
(c) Multifractal strength $-B_1$. 
\label{fig.4}}
\end{figure}

\begin{table}[h]
\centering
\caption{
Three periods selected for analysis and fitting parameters.
}
\hspace{-5mm}
\begin{tabular}{l|cc}
\hline
   Period   & $H_0$ & $a$   \\
\hline
   I: 20014/1/2-2022/1/2  & 0.1379(8)  & 2.93(11)   \\
   II: 2015/1/1-2023/1/1   & 0.1308(4)  & 3.02(6)   \\
   III: 2015/5/27-2023/5/27 & 0.1262(3)  & 3.15(6)   \\
\hline
\end{tabular}
\end{table}

\section{Conclusion}
We examine the finite sample effect on the HE using Bitcoin data and
find that the HE decreases as $\Delta$ increases.

We provide a simple two-parameter ansatz that obtains the HE at $\Delta \rightarrow 0$ and use ansatz to obtain the HE  $\sim 0.12-0.14$, indicating
the roughness of the RV time series.
We also find that the relative error of the HE for the five-minute RV is 
small ($1\%$); thus, we conclude that the five-minute RV can be used for the HE estimate without considering 
the finite sample effect.
Analyzing the five-minute RV time series, we
find that its multifractality is smaller than that of price-return 
and the strength of the multifractality varies over time.

Two sources of the multifractality are considered (i) non-linear time correlations 
and  (ii) shape of distribution~\citep{kantelhardt2002multifractal}.
The previous study suggests the multifractality originates in part from the shape of distribution\citep{takaishi2020rough}. 
It is important to understand which source dominates the multifractality in the time series 
when we model the RV time series. 
If the dominant source is the shape of distribution, one may need to
introduce the stochastic process from the distribution.
This might be similar to the scheme of the GARCH-type models 
in which to accommodate the fat-tailed return distribution,
non-normal distributions
are introduced to the innovations of the GARCH model\citep{bollerslev1987conditionally,Nelson1991Econ}.
If the non-linear time correlations dominate, one may need a dynamical process that generates the multifractality in time series.
Future researchers can investigate and clarify the origin of the multifractality in the RV time series.

A limitation of the study is that we examine only Bitcoin data. 
Future research could investigate 
whether our findings hold for other assets.
If we confirm the multifractality in the volatility time series,
it could serve as a new property that can guide to construct a reliable volatility modeling.

\section*{Acknowledgements}
The numerical calculations for this study were performed using the Yukawa Institute Computer Facility and facilities at the Institute of Statistical Mathematics.
This study was supported by
the Yu-cho Foundation (Grant-in-Aid for Research , 2024) and in part by JSPS KAKENHI, grant number JP21K01435.


\end{document}